\documentclass[12pt,preprint]{solarphysics}

\usepackage[hyperref,optionalrh,natbib]{spr-sola-addons} % For Solar Physics 

\usepackage{graphicx}        % For eps figures, newer & more powerfull
\usepackage{color}           % For color text: \color command
\usepackage{breakurl}        % For breaking URLs easily trough lines
\newcommand{\aap}{    {\it Astron. Astrophys.}}

\newcommand{\apjl}{   {\it Astrophys. J. Lett.}}

\begin{document}

\begin{article}

\begin{opening}

\title{The Helioseismic and Magnetic Imager (HMI) Vector Magnetic Field Pipeline: Optimization of the Spectral Line Inversion Code}

\author{R. Centeno$^{1}$ \sep
J. Schou$^{{2},{3}}$ \sep 
K. Hayashi$^{2}$ \sep 
A. Norton$^{2}$ \sep
 J.T. Hoeksema$^{2}$\sep  
 Y. Liu$^{2}$\sep 
 K.D. Leka$^{4}$\sep 
 G. Barnes$^{4}$
 }

\runningtitle{Modified VFISV}
\runningauthor{R. Centeno et al.}

	\institute{$^{1}$ High Altitude Observatory (NCAR), 3080 Center Green Dr., Boulder CO 80301\\
	$^{2}$ HEPL Solar Physics, Stanford University, CA 94305-4085\\
	$^{3}$ Max Planck Institute for Solar System Research, Max-Planck-Str. 2, 37191 Katlenburg-Lindau, Germany\\
	$^{4}$ NwRA, 3380 Mitchell Lane, Boulder, CO 80301\\
	}

\begin{abstract}
The Very Fast Inversion of the Stokes Vector (VFISV) is a Milne-Eddington spectral line inversion code used to determine the magnetic and thermodynamic parameters of the solar photosphere from observations of the Stokes vector in the 6173 \AA\ Fe {\sc i } line by the {\it Helioseismic and Magnetic Imager} (HMI) onboard the {\it Solar Dynamics Observatory} (SDO). We report on the modifications made to the original VFISV inversion code in order to optimize its operation within the HMI data pipeline and provide the smoothest solution in active regions. The changes either sped up the computation or reduced the frequency with which the algorithm failed to converge to a satisfactory solution. Additionally, coding bugs which were detected and fixed in the original VFISV release, are reported here. 

\end{abstract}

\end{opening}


\section{Introduction:  What is VFISV?}

The Very Fast Inversion of the Stokes Vector \citep[VFISV;][]{Borr11} is a spectral line inversion code tailored and optimized to invert the full disk spectro-polarimetric data of the {\it Helioseismic and Magnetic Imager} (HMI) instrument \citep{Scherrer2012, Schou2012} on board the {\it Solar Dynamics Observatory} \citep[SDO;][]{Pesnell2012}. HMI is a filtergram-type instrument that observes (with one camera) the Stokes {\it I}, {\it Q}, {\it U}, and {\it V} at six wavelength positions across the Fe {\sc i} 6173 \AA\ spectral line for the full disk of the Sun with 16 million pixels (4096$\times$4096 CCD) every 135 s.  All of the data pipeline procedures, and the spectral line inversion code in particular, have limited allowable runtimes in order to keep pace with the data flow rate and prevent a processing backlog. 

In the forward problem, VFISV solves the radiative transfer equation (RTE) for polarized light using the Milne-Eddington (ME) approximation to generate a set of Stokes profiles from a given model atmosphere \citep[][]{unno, rachkovsky}. The ME approximation assumes that all the parameters describing the atmosphere are constant along the line of sight (LOS) except for the source function, which varies linearly with optical depth. In addition, the generation of polarized radiation is formulated in the classical Zeeman effect regime. Traditionally, ME models applied to polarized RTE problems use up to eleven free parameters to describe the atmosphere in which the Stokes profiles are generated. There are five thermodynamical parameters: the line-to-continuum absorption ratio, $\eta_0$, the Doppler width, $\Delta\lambda_{\rm D}$, the damping parameter of the Voigt function, $a$,  and the components of the linearized source function,  $B_0$ and $B_1$, respectively.  The magnetic field vector is described by three variables, {\it i.e.} the magnetic field strength, $B$, its inclination with respect to the LOS, $\gamma$, and its azimuth in the plane perpendicular to the line of sight, $\psi$ (which is referenced to a column of pixels on the HMI CCD and increases counter-clockwise).
There are also two kinematic parameters:  the Doppler velocity, $v$, and the macroturbulent velocity, $v_{\rm mac}$. The former characterizes the macroscopic plasma speed, while the latter is generally used to model a combination of unresolved plasma velocity fields and instrument smearing effects -
it is usually expressed in the form of a convolution of the Stokes profiles with a gaussian function of width $v_{\rm mac}$.   VFISV makes explicit use of the measured HMI transmission filter profiles
in the spectral line synthesis and hence does not need to use $v_{\rm mac}$
to account for the instrumental broadening, while the other thermodynamic
parameters compensate for the unresolved velocities and any residual
instrumental broadening.
A standard additional geometrical parameter known as the filling factor, $\alpha$, quantifies the fraction of light within any given pixel that originates from a magnetized atmosphere. 

The optimization scheme of VFISV, based on a Levenberg-Marquardt (LM) minimization algorithm \citep[see][]{press}, takes a set of observed Stokes profiles and finds the model parameters that best describe the atmosphere in which they were generated. It achieves this by performing a non-linear minimization of a merit function, $\chi^2$, that measures the similarity between the observed and synthetic Stokes profiles in a least squares sense.  

VFISV is just one module among many in the data pipeline for the HMI instrument. It operates on Level1.5 data (the \verb|hmi_S.720s| data-series) from the 'vector' camera of HMI averaged every 720 s \citep[][]{hoeksema2014} and its output is fed to a disambiguation code \citep[see][]{barnes2014} that resolves the $180^\circ$ azimuth ambiguity of the magnetic field vector.
The original version of VFISV \citep{Borr11} was developed before the launch of SDO, which took place on 11 February 2010. The spectral line inversion code has been further optimized since HMI data became available online. The purpose of this paper is to describe the changes implemented in the code and its output since the launch of SDO.

The current version of the code is referred to as \verb|fd10|, which began
as a number (in this case 10) assigned to identify tests of various versions
of the code on the full-disk (FD) data.
The data series produced by VFISV \verb|fd10| in the HMI pipeline processing is called \verb|hmi.ME_720s_fd10| and is available through JSOC\footnote{http://jsoc.stanford.edu/} (Joint Science Operations Center). Its name reflects the fact that it is a product of the Milne-Eddington inversion (ME) every 720 s with the \verb|fd10| version of the VFISV code.
 
This manuscript is organized as follows. Section \ref{section:chi2-space} reports on the changes in the code that alter the $\chi^2$-space and hence determine which families of solutions are preferred over others, whilst Section \ref{section:speed} describes all of the procedures and alterations that have a direct impact on the speed performance of the inversion but not on the final solution found by the algorithm. Section \ref{section:conclusions} sums up the improvements in speed and performance of the code and some bug fixes are reported in the Appendix.

\section{Changes in the $\chi^2$ Space} \label{section:chi2-space}

The changes implemented in the VFISV code since 2010 that are discussed in this section have an impact on the position of the minimum of the merit function, $\chi^2$, that the algorithm is designed to minimize. Changes in the depth of the minima or the steepness of the function along a given parameter have an effect on the path that the algorithm takes to find a solution and on the solution itself.  Subject to the limits set on parameters as described in Section 2.3 and Table 1,
the objective of the code is always to find the global minimum.

\subsection{Weighting of the Stokes Profiles}

The inversion of the HMI Stokes vector is based on a non-linear least-squares minimization of a merit function, $\chi^2$, that measures the difference between the observed and synthetic Stokes profiles ({\it i.e.} the goodness of the fit). This merit function is defined as:

\begin{equation}
\chi^2 = \frac{1}{F} \sum_S W_s^2 \sum_{\lambda} \frac{({\rm OBS}_{S}(\lambda)
- {\rm SYN}_{S}(\lambda, \mathbb M))^2}{\sigma_S^2}
\label{eq:chi2}
\end{equation}

\noindent where $F$ is the number of degrees of freedom ({\it i.e.} the number of data points minus number of free parameters in the model), OBS and SYN refer to
the observed and synthetic Stokes profiles, respectively, and $\mathbb M$ represents the model atmosphere. The index $S$ denotes the 4 Stokes parameters and $\lambda$ denotes the wavelength positions. $\sigma_S$ represents the photon noise, which is a function of the intensity and the polarization state\footnote{The expressions used to calculate the noise  for the Stokes parameters come from accounting for the gain, the exposure time, the temporal averaging and the number of filtergrams that go into calculating each Stokes parameter. The latter results in different values of $\sigma$ for the intensity and the polarization: $\sigma_I=0.118\cdot\sqrt{I_{\rm C}}$ and
$\sigma_{\rm Q,U,V}=0.204\cdot \sqrt{I_{\rm C}}$. This arises as a consequence of the chosen polarization modulation scheme, described in \cite{hoeksema2014}}.  This is needed because a different number of filtergrams go into constructing the intensity and the polarization images. However, the intrinsic wavelength dependence of $\sigma_S$ is not accounted for in \verb|fd10|, rendering it a function of the continuum intensity and the Stokes parameter only. The weights, $W_S$, are a set of four values (one for each Stokes parameter), whose purpose is to emphasize or de-emphasize the relative importance of a given Stokes parameter with respect to Stokes {\it I} in its contribution to $\chi^2$.

The amplitude of the Stokes {\it V} signal is usually $10^{-1} - 10^{-2} \times I_{{\rm C}}$ (where $I_{{\rm C}}$ is the continuum intensity) while Stokes {\it Q} and {\it U} are typically one order of magnitude smaller. For this reason, the contribution of the {\it Q}, {\it U}, and {\it V} profiles to $\chi^2$ is often almost negligible compared to that of Stokes {\it I}. However, Stokes {\it Q}, {\it U}, and {\it V} carry most of the information about the magnetic field. While Stokes {\it I} is only sensitive to the magnetic field strength in strong field regions, it is independent of azimuth and only slightly dependent on the inclination with respect to the LOS.

There are a number of factors that can negatively impact the quality of the fit
of the Stokes vector in an inversion. Flatfielding errors or inaccuracies in our
knowledge of the instruments's filter profiles could lead to
systematic "missfits" of the Stokes parameters. In the case of HMI, the
latter is likely to have a larger effect than the former. However, a more important
source of error is that the physical model, ${\mathbb M}$, used to describe the
atmosphere is not sufficiently realistic. 
Whatever the reason, an error in the fitting of the Stokes profiles will scale with their
amplitude in the expression of $\chi^2$, and because Stokes {\it I} is
orders of magnitude bigger than {\it Q}, {\it U}, and {\it V}, its contribution will override all of
the others. In other words, a slight missfit of Stokes {\it I}  will always yield a non-zero difference of ${\rm OBS}_{I}(\lambda) - {\rm SYN}_I(\lambda, {\mathbb M})$ in the expression of $\chi^2$. Due to the much larger amplitude of Stokes {\it I} compared to Stokes {\it Q}, {\it U}, and {\it V}, this term can easily mask equivalent differences for the polarization profiles ({\it i.e.} ${\rm OBS}_{S}(\lambda) - {\rm SYN}_S(\lambda, {\mathbb M})$, with $S = Q, U, V$).
In order to de-emphasize Stokes {\it I} relative to {\it Q}, {\it U}, and {\it V}, we use a set of weights, determined empirically, that balances out the amplitude of the different Stokes signals in the expression of $\chi^2$. 

Because the expression of $\chi^2$ acknowledges dependence of the photon noise on the Stokes parameter ($\sigma_I \ne \sigma_{Q,U,V}$ and $\sigma_Q=\sigma_U=\sigma_V$), it is useful to define the concept of {\em effective weights} as $W^{\rm eff}_S = W_S / \sigma_S$. 
The weight scheme currently adopted for VFISV is $W_S \approx$ [1,5,5,3.5] (or, equivalently, $W^{\rm eff}_S \approx $[1,3,3,2]) for [{\it I,Q,U,V}], respectively (note that these weights are then squared in Equation (\ref{eq:chi2})). The effect of the weights is to change the shape of the $\chi^2$ function and how steep the gradients along the different model parameters are. 

\begin{figure}[!t]
\centerline{
\includegraphics[width=\textwidth,bb=54 320 550 800, clip=]{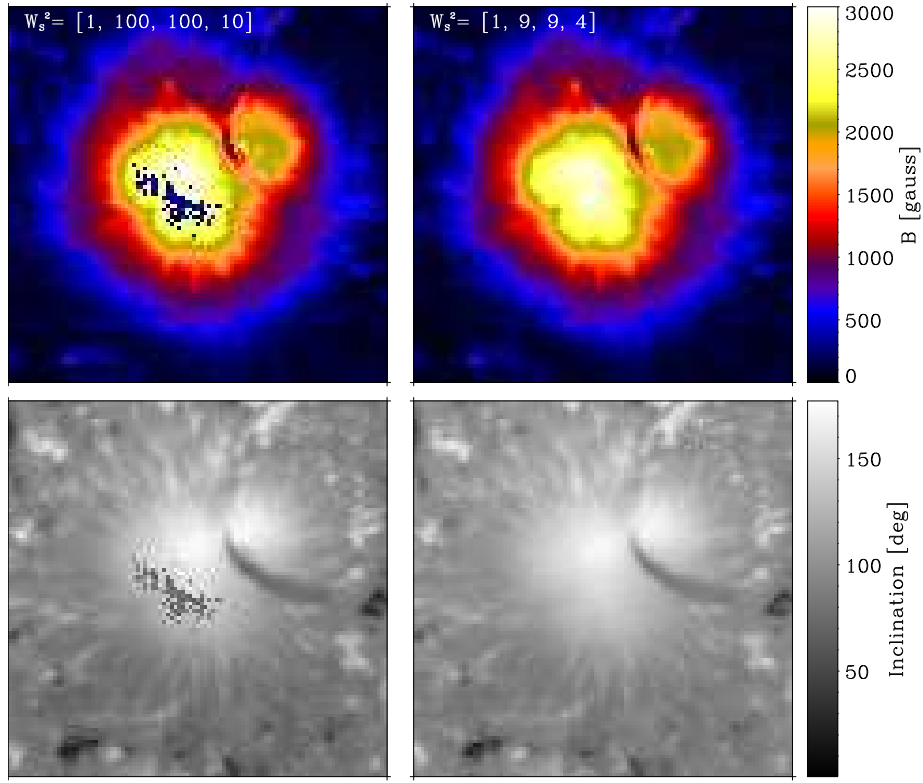}}
\caption{Magnetic field strength (top) and inclination (bottom) retrieved from the inversion of an AR using two different set of weights, that of the {\it Hinode}/SOT-SP MERLIN inversion code $W_S^2=[1,100,100,10]$ (left) and the one used in the fd10 version of the HMI data pipeline, $W_S^2=[1,9,9,4]$ (right). The inhomogeneity of the retrieved magnetic field in the umbra when using the MERLIN weights becomes apparent.}
\label{fig:weights}
\end{figure}

The Sun hosts a range of magnetic activity, producing an array of polarization signatures in spectral lines.  It is common to weight the Stokes profiles with a custom set of weights for each pixel, in order to
balance out the amplitudes of the four Stokes profiles. In the original version of VFISV, \citet{Borr11} proposed $W_{Q,U,V}=1$ and $W_{I} < 1$, where the weight for Stokes {\it I} was lowered more or less depending on the value of the continuum intensity for the pixel under consideration.  However, for the sake of homogeneity, VFISV \verb|fd10| has been altered so that the weighting scheme is the same for each and every pixel. 
\noindent Other instrument teams have implemented inversion codes with constant values for $W_S$ so that $\chi^2$ values can be more easily compared. For instance, MERLIN\footnote{https://www.csac.hao.ucar.edu/} (the spectral line inversion code for the {\it Hinode}/SOT-SP data pipeline, which is direct heritage of the {\it Advanced Stokes Polarimeter} inversion code \citep{ASPcode}), operates with constant values $W_S^2$= $[1,100,100,10]$ (or, equivalently, $W_S^{\rm eff}$= $[1,10,10,\sqrt{10}]$) for [{\it I,Q,U,V}] \citep{Lites07}. This ratio for the weighting of the Stokes profiles is standard for high spectral resolution data from spectrograph instruments. However, the spectral smearing induced by the width of the HMI transmission filter profiles produces qualitatively different spectral line shapes and depths that call for custom $W_S$ values. Figure \ref{fig:weights} shows the differences in the magnetic field strength (top) and inclination with respect to the LOS (bottom), retrieved from the inversion of an AR observed by HMI using two different sets of weights, namely $W^{\rm eff}_S=[1,10,10,\sqrt{10}]$ in the left panels, and $W^{\rm eff}_S=[1,3,3,2]$ in the right panels. The MERLIN weighting scheme (left panels) does not work for the HMI data, and the inhomogeneity in the retrieved magnetic field parameters becomes apparent in strong field regions.
The chosen set of weights for the HMI data, $W^{\rm eff}_S=[1,3,3,2]$, proved to yield the smoothest solution inside ARs.

The HMI vector team adopted the policy of uniform weights throughout the solar disk. The final choice of weights was selected to optimize the inversion results in active regions (ARs) with the aim of achieving smooth solutions for the magnetic field inside the umbrae and avoid discontinueties that would arise from changing weights. As a side effect, the weighting scheme is far from optimal for quiet Sun, where the adopted set of weights leads to artifacts in the solution for the inferred magnetic fields - it exacerbates the ``horizontal field effect'' in which the inversion of pixels that have near-to-pure noise polarization signals will deliver a seemingly strong purely transverse magnetic field vector \citep{borrero_kobel}. 

\subsection{Reducing the Number of Free Parameters}\label{section:freeparams}

Inverting the HMI data is challenging due to a combination of moderate spectral resolution and very coarse spectral sampling.  Since there are only six wavelength sampling positions across the 6173 \AA\ Fe {\sc i} spectral line, we have a relatively small number of observations with which to constrain a model with up to eleven free parameters.   Observing a single spectral line instead of a line pair (such as the 6302 and 6301 \AA\ Fe {\sc i} line pair) means we further decrease our ability to constrain the physical system.  Using two spectral lines, as {\it Hinode}/SOT-SP does, takes advantage of the different line formation physics (as evident in the Land\'e factor) to distinguish between parameters otherwise redundant. In general, there are degeneracies in the thermodynamical parameters of the model when carrying out the inversion.  One approach to improving the performance of VFISV was to reduce the number of free parameters and thus decrease its degrees of freedom.  

Early tests (\cite{Borr11}, Figure 14) showed that it was possible to fix the damping to a constant value and to ignore the macroturbulence without a loss in accuracy in the determination of the magnetic field vector due to the sparse sampling of the spectral line. Setting the damping to a fixed value reduced the computing time as well as the degeneracy of the results. Originally, the damping parameter was set to a constant value of $a=1.0$, but further analysis showed that when solved for as a free parameter, the average value was closer to $a=0.5$ for the height of formation of the 6173 \AA\ line.  Therefore, the damping parameter was set to a constant value of $a=0.5$ for the production of \verb|fd10| data.  Furthermore, the magnetic flux, $\alpha \times B$, is well-constrained in the fit, but the individual values of $\alpha$ and $B$ are not.  For this reason, the filling factor is set to a constant value of $\alpha=1$.   A limitation introduced by assuming $\alpha=1$ is that the returned magnetic field strength value is in fact an area-averaged quantity \citep{Lites+Skumanich1990}.  In areas with unresolved magnetic structures such as plage, where averaged values of $\alpha$ are approximately 0.15 \citep{Martinez97}, the \verb|fd10| field strength values will be significantly lower than the intrinsic field strength that would result from an inversion in which the filling factor were allowed to be less than one.  In addition, the inclination angle will be biased towards the line of sight \citep{LekaBarnes12}.  

\subsection{Limits on Variables}

\begin{table}[ht]
\caption{Upper and lower limits on the model parameters in VFISV.} % title of Table
%centering % used for centering table
\begin{tabular}{l c c c c} % centered columns (r columns)
\hline\hline %inserts double horizontal lines
Parameter & Symbol & Units & Lower limit & Upper limit \\ [0.5ex] % inserts table
%heading
\hline % inserts single horizontal line
Doppler width & $\Delta\lambda_{\rm D}$ &  m\AA & 1 & 500  \\ 
Line-to-continuum ratio & $\eta_0$ & dimensionless & 1 & 1000 \\
Doppler velocity & $v$ & cm\,s$^{-1}$ & $-7\cdot10^{5}$ & $7\cdot10^{5}$ \\
Source function& $B_0$ & DN\,s$^{-1}$ & 0.15 $\times I_C$  & 1.2 $\times I_C$  \\ 

Source function gradient& $B_1$ & DN\,s$^{-1}$ & 0.15 $\times I_C$ & 1.2 $\times I_C$ \\
Field strength& $B$ &gauss & 5 & 5000 \\
Inclination & $\gamma$ &$^{\circ}$ & 0 & 180  \\
Azimuth & $\psi$ & $^{\circ}$ & 0 & 180 \\ 
\hline %inserts single line
\end{tabular}
\label{table:limits} % is used to refer this table in the text
\end{table}

%Occasionally, the inversion algorithm fails to find an adequate location in the parameter space that would place it on a path leading to the solution. Instead, in its attempts to find a better starting point, it ends up going too far out in the parameter space, i.e., too far away from any reasonable physical scenario for the formation of the spectral line. The result of this is that the algorithm is unable to find the direction that points towards the solution (i.e. the global minimum of $\chi^2$), helplessly diverging away from it until it hits the maximum number of iterations that it is allowed.
Occasionally, the iteration algorithm will find a location with a
low $\chi^2$ value, but unphysical parameters.
In order to constrain the parameter search to a reasonable set of values, VFISV
sets lower and upper limits to each of the physical parameters of the ME model atmosphere (see Table \ref{table:limits}). These limits must be physically meaningful, at least in those cases where the ME parameter has a direct correspondence with a real physical magnitude. For instance, the magnetic field strength is not expected to exceed the 5000 G (gauss) limit, nor can it be less than 0 G. Even if an AR did harbor magnetic fields stronger than 5000 G, the dynamic range of the HMI instrument only reaches up to $\approx3200$ G when accounting for the large range of spacecraft velocity \citep{Liu}, so the 5000 gauss (G) upper threshold covers the measurable range. Occasionally, when the observed Stokes profiles originate in the midst of very high field strengths ({\it i.e.} very dark places inside large sunspot umbrae) they challenge the dynamic range of the instrument and the inversion algorithm is unable to converge to the solution, hitting the hard limit set at 5000 G. 

The limits on the LOS velocity, $v$, were determined taking into account typical plasma motions at the photosphere, the effects of solar rotation, the satellite's orbit around the Earth (which, alone, induces Doppler shifts of up to $\pm 3.5$ km\,s$^{-1}$) and the Earth's orbit around the Sun. The inclination with respect to the LOS can only vary between 0 and 180$^{\circ}$, and although the azimuth spans the entire 0-360$^{\circ}$ range, there is an inherent 180$^{\circ}$ ambiguity when determining it from the Stokes profiles alone. The terms of the source function should add up to the continuum intensity, thus, neither of them is allowed to exceed $1.2 \times I_{\rm C}$. There is a strong degeneracy between these two parameters, hence an arbitrary lower limit was set, just to prevent negative values.  The constraints on the rest of the variables were empirically derived and are justifiable from a $\chi^2$ minimization perspective.

\subsection{Regularization of $\chi^2$}\label{section:regularization}

In the case of VFISV \verb|fd10|, $\chi^2$ represents a hypersurface of eight independent variables ({\it i.e.} the eight free parameters of the model atmosphere). In reality, due to limitations of the model and the data, these parameters are not all independent. Some degree of degeneracy exists among pairs and even combinations of several of them. This typically causes the $\chi^2$ function to host narrow flat valleys where the LM algorithm cannot find the solution. One clear example of this is the degeneracy between the magnetic filling factor and the field strength in the weak field regime, {\it i.e.}, the Stokes profiles caused by a relatively-strong field with a small filling factor and those caused by a weak field with a large filling factor, are indistinguishable \citep[see, for instance][]{marian2006}.  In the \verb|fd10| version of the HMI pipeline inversion code the filling factor is set to a fixed value of 1, as described in Section \ref{section:freeparams}.   Another example of degeneracy between parameters is described in Section \ref{sec:var_change}.

A related problem is the presence of multiple minima in the $\chi^2$ function.
When there is a tendency for $\chi^2$ to host two or more minima of similar depth in different parts of the parameter space, the algorithm will lead to each of these solutions with a certain degree of probability.
Extensive tests of the inversion code on HMI data revealed a degree of degeneracy between $\eta_0$ and $B$. This particular degeneracy presented itself as a consistent double minima in the $\chi^2$ function, with a bi-modal distribution of $\eta_0$ values. There would typically be two solutions with magnetic field strengths that differed by 100-300 G, one associated with a small value of $\eta_0$ and another one associated with a large value.  

\begin{figure}[!t]
\centerline{
\includegraphics[width=\textwidth,bb=54 340 550 800, clip=]{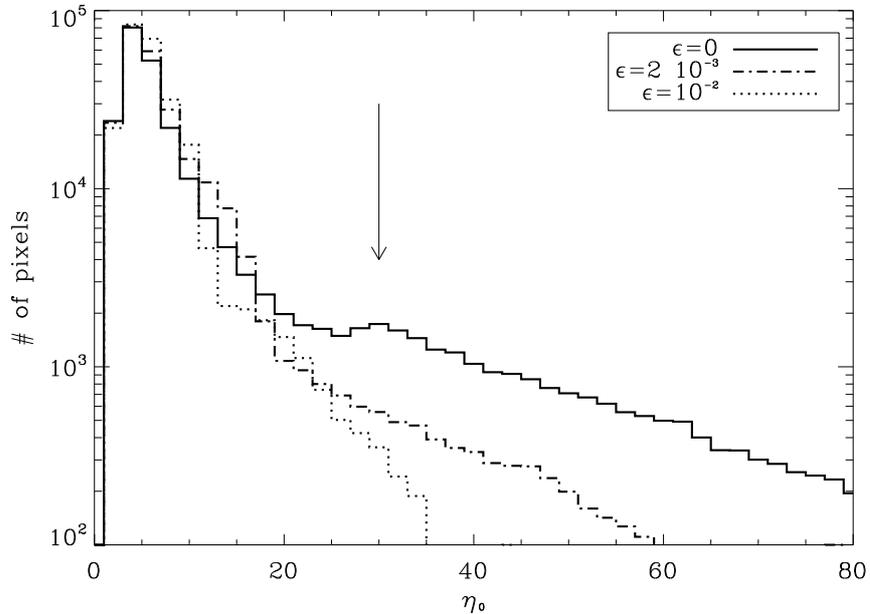}}
\caption{Effects of different regularization terms on the histogram of $\eta_0$ values. The solid line corresponds to no regularization, where the arrow points at the second hump in the bi-modal distribution of $\eta_0$. The dashed line corresponds to the regularization implemented in the VFISV fd10 pipeline (with $\epsilon=2\times10^{-3}$), while the dotted line represents the histogram of $\eta_0$ for a larger value of $\epsilon= 10^{-2}$.}
\label{regularization}
\end{figure}

There is no physical reason to expect a bimodal distribution of $\eta_0$ values. The degeneracy between {\it B} and $\eta_0$ is largely a consequence of the limited spectral resolution of the HMI data. Milne-Eddington inversion results of {\it Hinode}/SOT-SP data (which have a spectral sampling of $\approx 21$ m\AA\ per pixel) largely favor lower values of $\eta_0$. For this reason, high $\eta_0$ values were deemed unphysical, and an empirically determined regularization term was added to $\chi^2$  in order to get rid of them:

\begin{equation}
\chi^2_{\rm new} =  \chi^2_{\rm old} + \epsilon (\eta_0 - C)^2
\label{eq:reg} 
\end{equation}

\noindent where $\epsilon=0.002$ and $C=5$. This term penalizes high values of $\eta_0$ in the merit function and reduces the number of pixels that are susceptible to a double minima behavior. Figure \ref{regularization} shows the effects that different values of $\epsilon$ have on the histogram of $\eta_0$. The solid line corresponds to the case with no regularization term in $\chi^2$, in which case a bi-modal distribution of $\eta_0$ becomes apparent (the arrow points to the secondary hump in the histogram, for $\eta_0 \approx 30$). The typical distribution of $\eta_0$ values for VFISV \verb|fd10| is shown by the dashed line. With the constraint imposed by the regularization term, the secondary hump disappears, and the inversion favors lower values of $\eta_0$.

\section{Speed Optimization}\label{section:speed}

This section reports on modifications of the VFISV code aimed at speeding up the inversion without altering the shape of the merit function. 

\subsection{Explicit and Non-explicit Calculation of the Line Profile}

\begin{figure}[!t]
\centerline{\includegraphics[angle=0, scale=0.6]{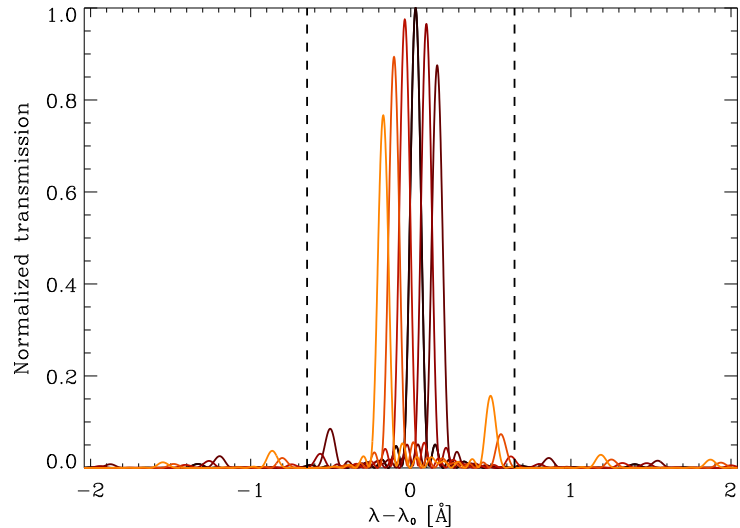}}
\caption{HMI transmission filter profiles as a function of wavelength, calculated for a range of [-2, 2] \AA\ from the center of the HMI spectral line. The vertical dashed lines delimit the range in which the forward modeling of the spectral line is calculated explicitly in the inversion code.}
\label{fig:hmi_filters}
\end{figure}

Figure \ref{fig:hmi_filters} shows the six HMI transmission filters as a function of wavelength
($\lambda_0$ corresponding to the central wavelength of the Fe {\sc i} 6173 \AA\ line). Each filter profile is characterized by a distinctive primary lobe and a number of smaller side lobes throughout the wavelength domain.

\begin{figure}[!t]
\centerline{
\includegraphics[angle=0, scale=0.22]{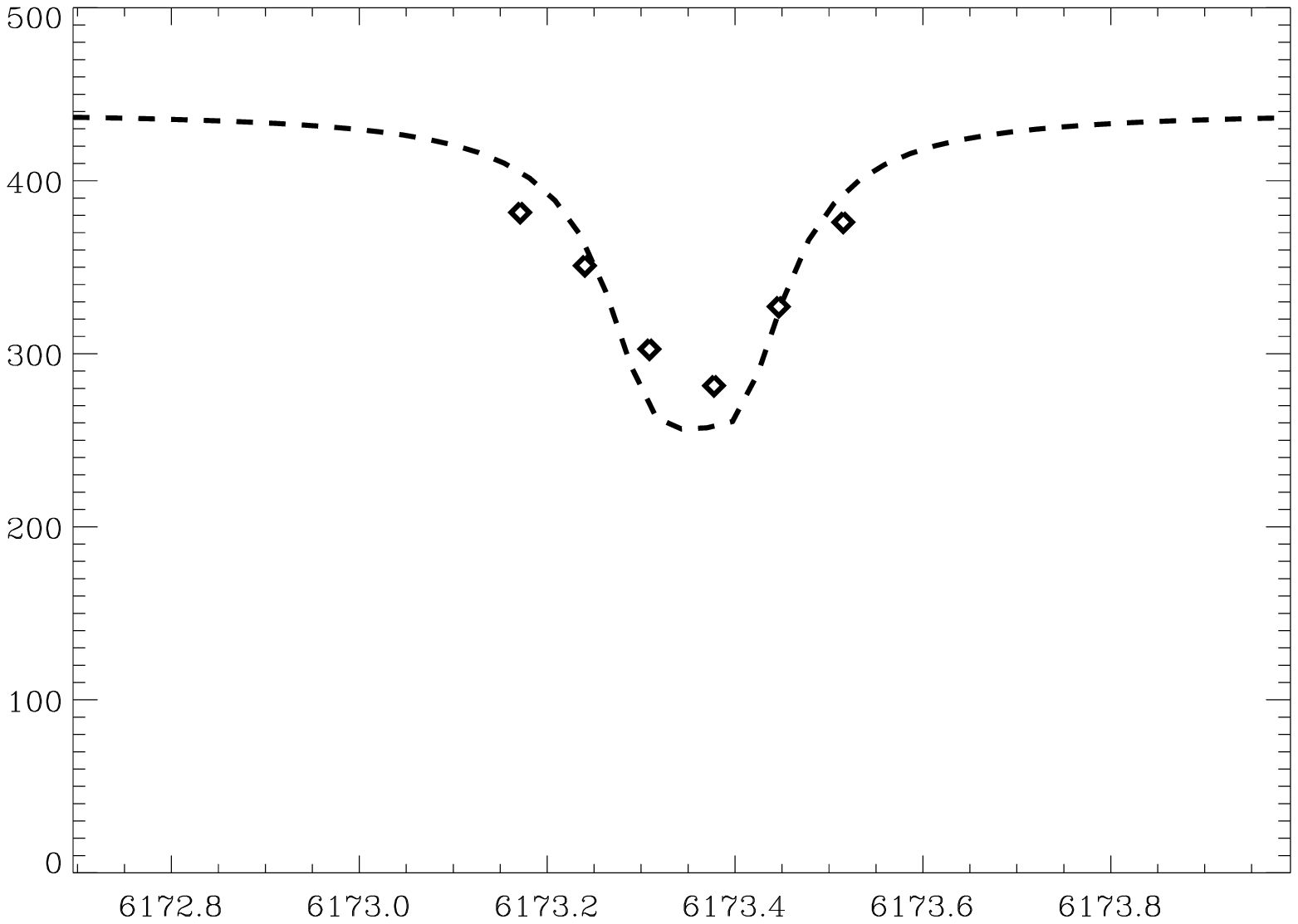}
\includegraphics[angle=0, scale=0.22]{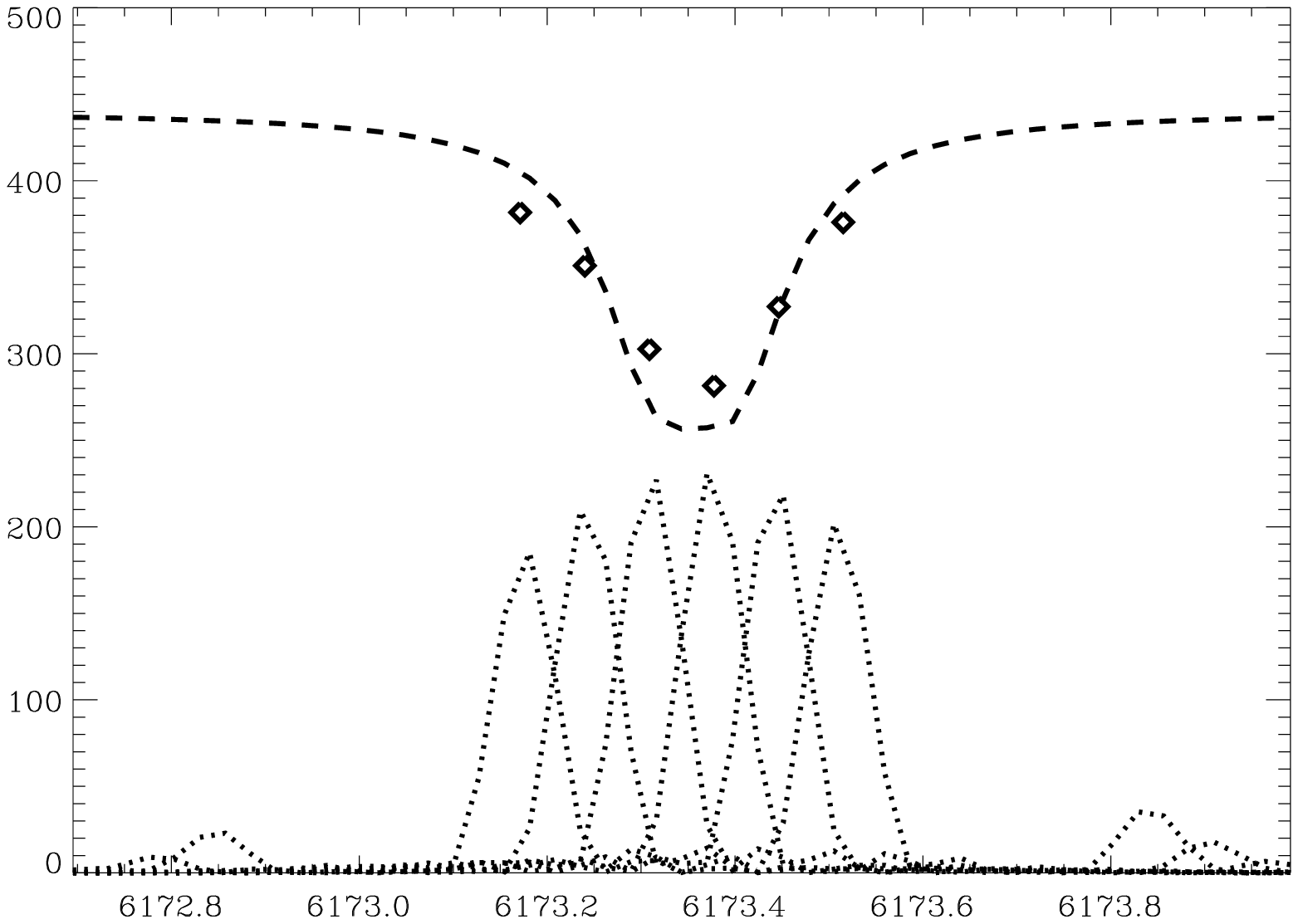}
\includegraphics[angle=0, scale=0.22]{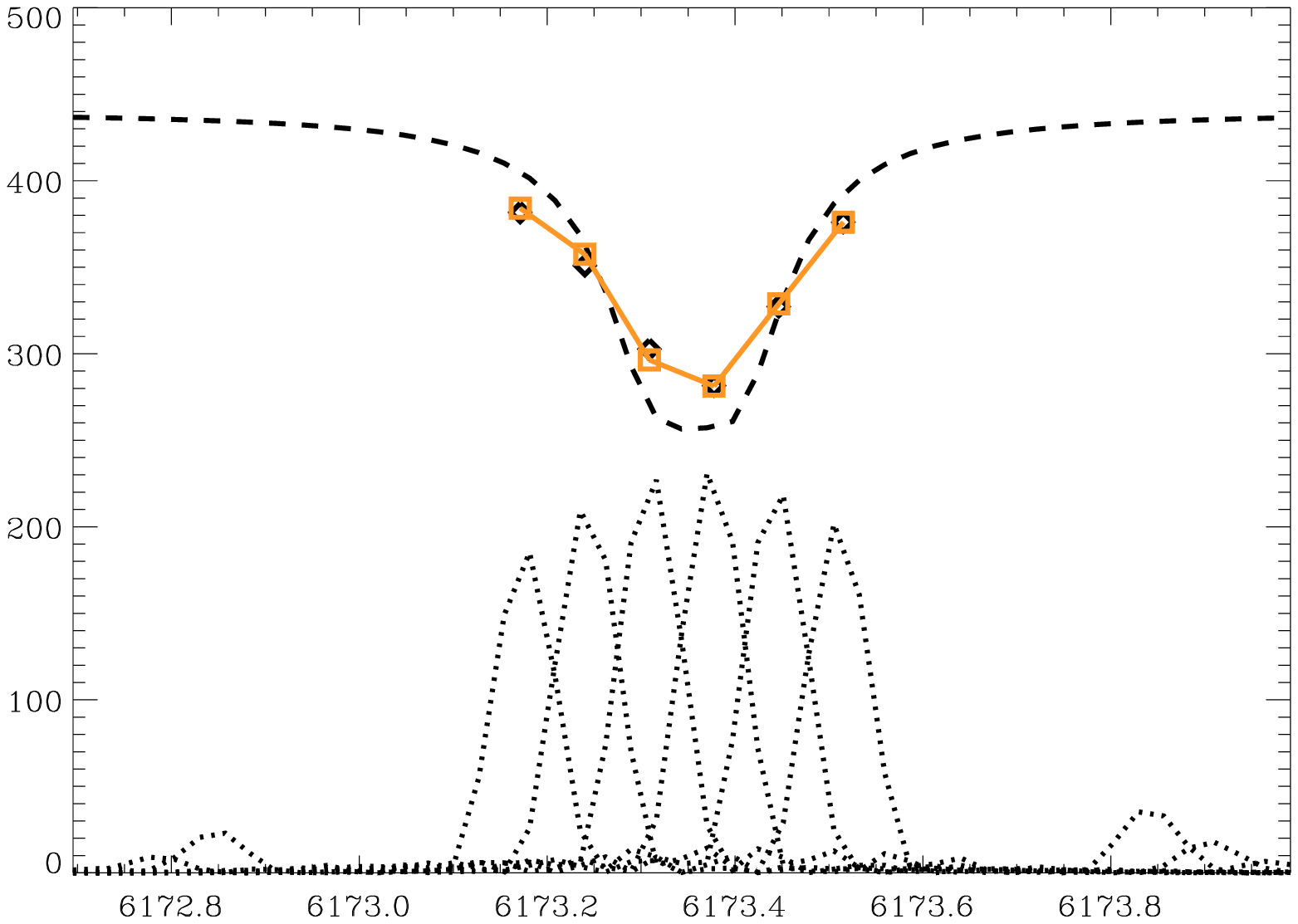}}
\caption{ Sequence of images showing, schematically, how VFISV incorporates the HMI filter profiles into the spectral line modeling. From left to right: first panel shows the observed Stokes {\it I} profile (diamonds) and the synthetic spectral line (dashed). The second panel includes the HMI filter profiles for the observed pixel (dotted lines). The third panel superposes the synthetic filtered Stokes {\it I} profile (in orange), obtained as a result of integrating the synthetic spectral line under each of the HMI filter profiles.}
\label{fig:hmi_sampling}
\end{figure}

In its forward modeling module, VFISV synthesizes the Stokes profiles of the Fe {\sc i} 6173 \AA\ spectral line with a relatively high spectral resolution. It then integrates in wavelength \citep[see Equation (3) in][]{Borr07} the products of the spectral line with the six HMI filters to generate HMI-like Stokes profiles. The result of this process, schematically shown in Figure \ref{fig:hmi_sampling}, is a set of synthetic filtered spectral profiles that are comparable to the data.

The wavelength range and sampling used to synthesize the HMI spectral line and the filter profiles are customizable parameters in VFISV.  The larger the range and finer the sampling, the more accurate the modeling of the spectral line will be. When synthesizing the Fe {\sc i} spectral line and applying the transmission profiles, it is necessary to account for the secondary lobes of the filters. Not doing so will lead to an underestimate, of up to 7\%, of the amount of light that passes through the filters \citep[see][for details on the calibration procedure for the HMI transmission profiles]{couvidat12}. Hence, it is important to extend the calculation as far into the wings of the line as practical.
However, the wider the spectral range, the larger the number of wavelength points in the spectral line synthesis calculation. This becomes rather computationally expensive. It is possible to reduce the number of points by lowering the spectral sampling of the synthesis, but this can only be done in detriment of the results. 

A compromise has to be reached between increasing the spectral range and reducing the sampling.
An alternative approach is to compute the contribution of the filter profiles far out in the continuum without doing the forward modeling of the spectral line in this region and without compromising the wavelength sampling. Figure \ref{fig:hmi_filters} shows the HMI transmission profiles. The dashed lines represent the lower and upper wavelength boundaries within which the intensity profile of the HMI spectral line is always contained (in the absence of Zeeman splitting), when accounting for the Doppler shifts induced by the dynamic range of the satellite, the solar rotation, and typical photospheric motions\footnote{Assuming $\pm$3.5\,km\,s$^{-1}$ line-of-sight velocity between the Sun and spacecraft, $\pm$2\,km\,s$^{-1}$ of solar rotation and $\pm$1.5\,km\,s$^{-1}$ of photospheric plasma motions, the possible induced Doppler shift can amount up to $\pm$0.144 \AA. Under typical photospheric conditions, the spectral line is less than 1\AA\ wide ({\it i.e.}, less than 0.5\AA\ at each side of the position of the core), so the entire profile will always be contained in a $\pm0.65$ \AA\ range around the central wavelength, in the absence of a magnetic field that would induce a Zeeman splitting.}. In \verb|fd10|, VFISV does the forward modeling only in this {\em inner} range ({\it i.e.} the spectral line is explicitly synthesized and the filter profiles are directly applied). In the {\em outer} range that goes farther into the wings the contribution of each filter profile is merely integrated, multiplied by the continuum intensity derived from the inversion and added to the synthetic Stokes {\it I} (this does not affect Stokes {\it Q}, {\it U}, and {\it V} because the continuum polarization is negligible). This is a reasonable approach when we work under the assumption that there is no spectral feature in the {\em outer} range. This strategy allows us to account for the contribution of the secondary and tertiary lobes of the instrument's transmission profiles without having to do the explicit calculation of the spectral line in this outer range. 

\begin{figure}[t]
\centerline{
\includegraphics[width=\textwidth,bb=0 300 620 860, clip=]{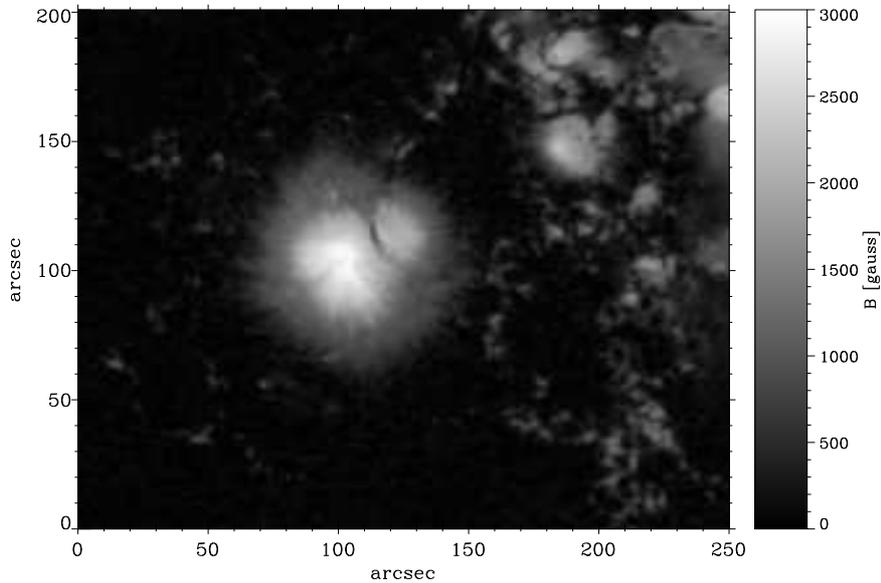}}
\caption{Magnetic field retrieved from an inversion where the full $\pm 2$ \AA\ wavelength range is considered explicitly.}
\label{fig:filter_hacking1}
\end{figure}

\begin{figure}[t]
\centerline{
\includegraphics[width=0.49\textwidth,bb=0 300 510 800, clip=]{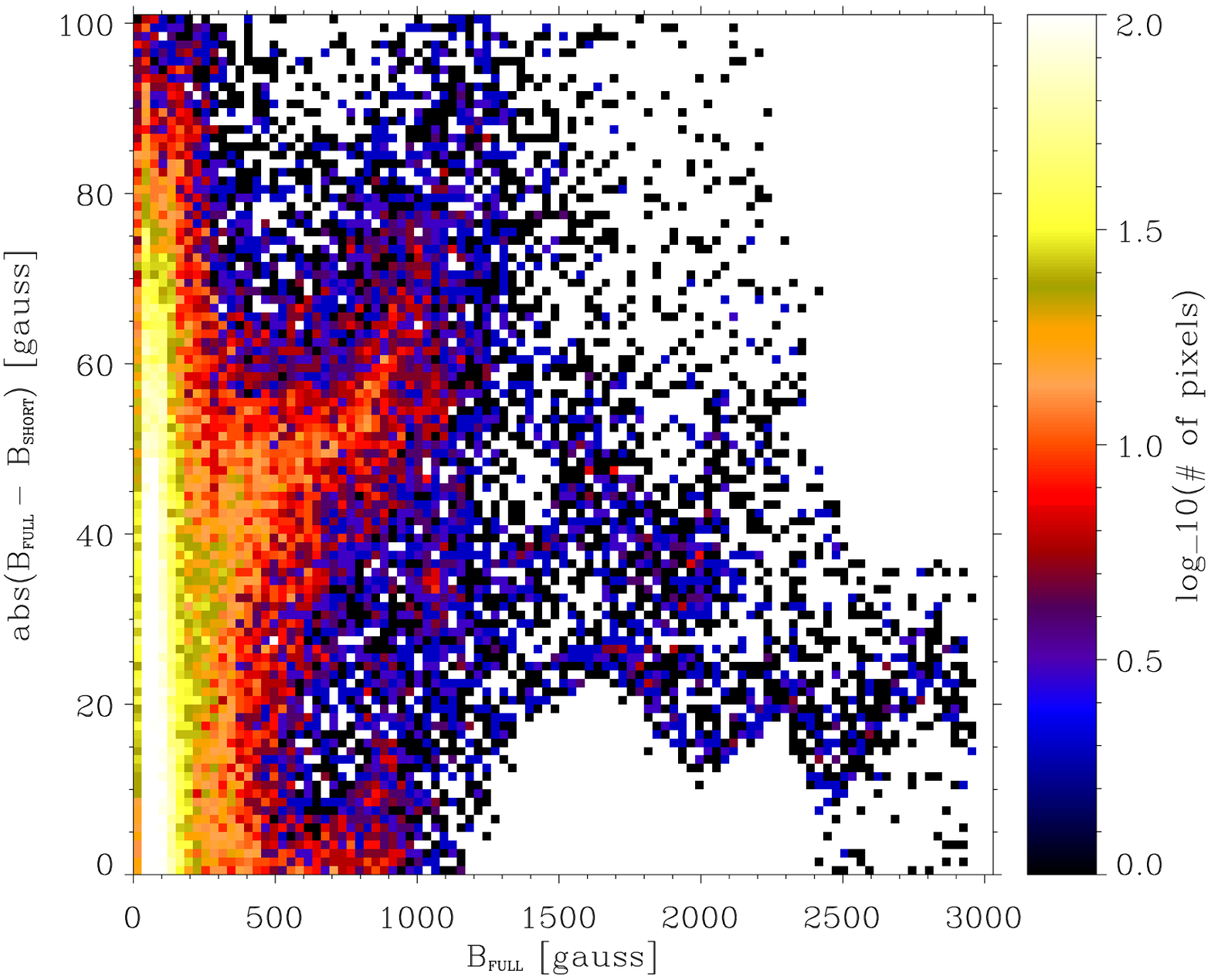}
\includegraphics[width=0.49\textwidth,bb=0 300 510 800, clip=]{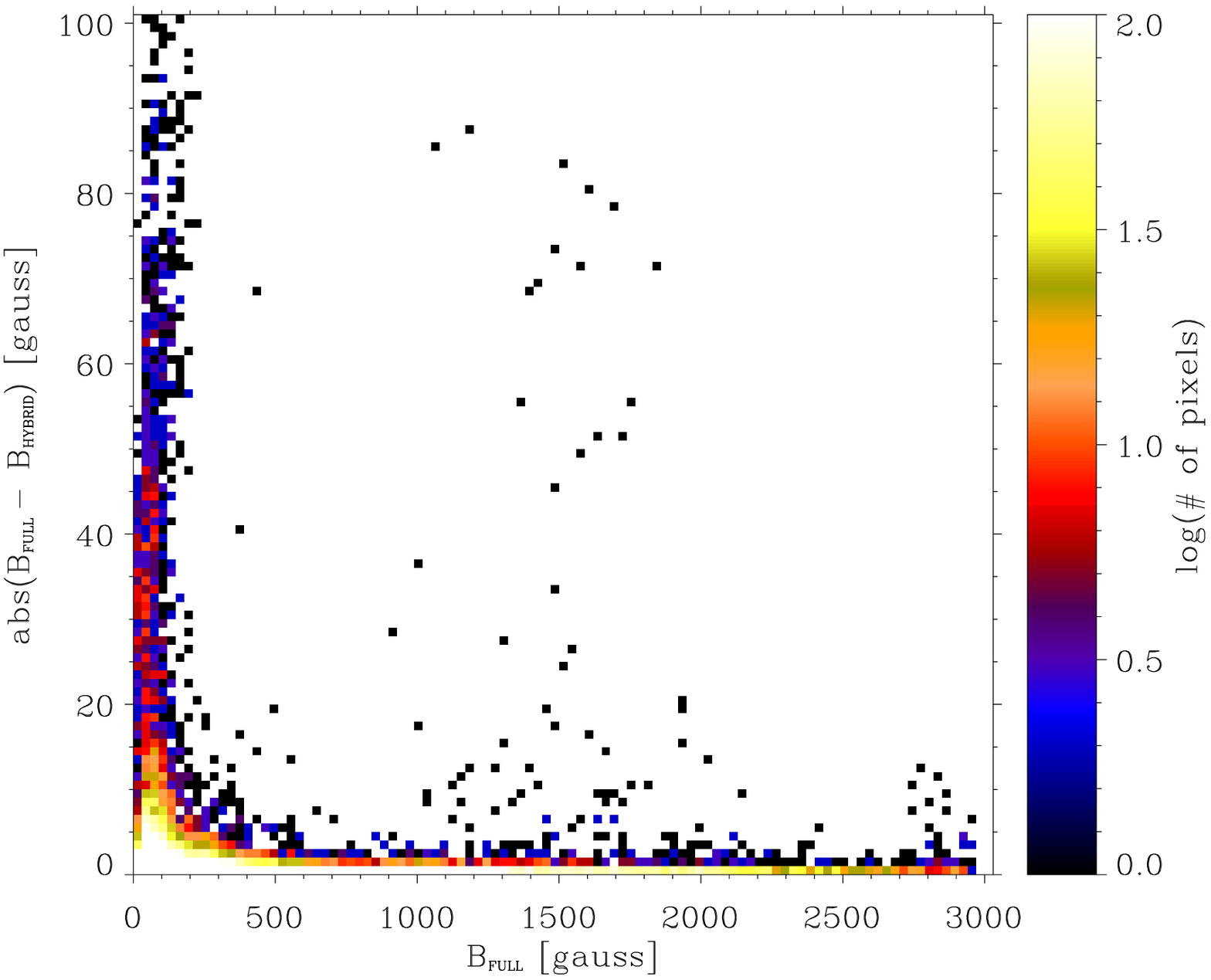}}
\caption{Effects of not including the secondary lobes of the HMI transmission profiles in the forward modeling. The left panel shows a density scatter plot of the difference between the magnetic field strength obtained from the full $\pm 2$\AA\ explicit calculation and the result of the inversion using only the {\em inner} $\pm 0.65$ \AA\ range. The right panel shows a scatter plot of the difference between the magnetic field of the full range explicit calculation and the result obtained using the hybrid approach. The differences in the retrieved magnetic field are visibly reduced when using the hybrid calculation.} \label{fig:filter_hacking2}
\end{figure}

A series of tests revealed that the extended spectral range should at least cover $\pm 2$ \AA\ from the line core, while the explicit calculation only needs to be carried out in the $\pm 0.65$ \AA\ range.
Figure \ref{fig:filter_hacking1} shows the magnetic field retrieved from the inversion of an AR using the explicit forward modeling in the full $\pm 2$\AA\ spectral range. A comparison between this and the results of an inversion using only the {\em inner} $\pm 0.65$ \AA\ range, exposes the problems derived from not accounting for a large enough wavelength range in the synthesis of the spectral line (left panel of Figure  \ref{fig:filter_hacking2}). Here, the difference between the retrieved magnetic fields from both inversions is shown in the form of a scatter plot. Inside the sunspot ({\it i.e.} for large fields), the relative difference in the magnetic field strengths amounts as much as 5\%. It is clear that the full wavelength range has to be considered in the calculation for the sake of accuracy. The right panel of Figure  \ref{fig:filter_hacking2} shows the difference in the inferred magnetic field strengths between the case of the explicit full range calculation and the hybrid approach described above (in which the explicit calculation is done in the {\em inner} $\pm 0.65$ \AA\ range, and the remaining contribution of the spectrum is calculated implicitly). In this case, the number of pixels for which the difference in retrieved magnetic fields is large has been significantly reduced.

Since this hybrid approach allows us to limit the explicit calculation to a third of the total necessary spectral range, this technique results in a speed up of a full disk inversion by a factor of $\approx 2.75$.

\subsection{Variable Change}
\label{sec:var_change}

\begin{figure}[!t]
\centerline{
\includegraphics[angle =90,width=0.48\textwidth,bb=50 50  600 700, clip=]{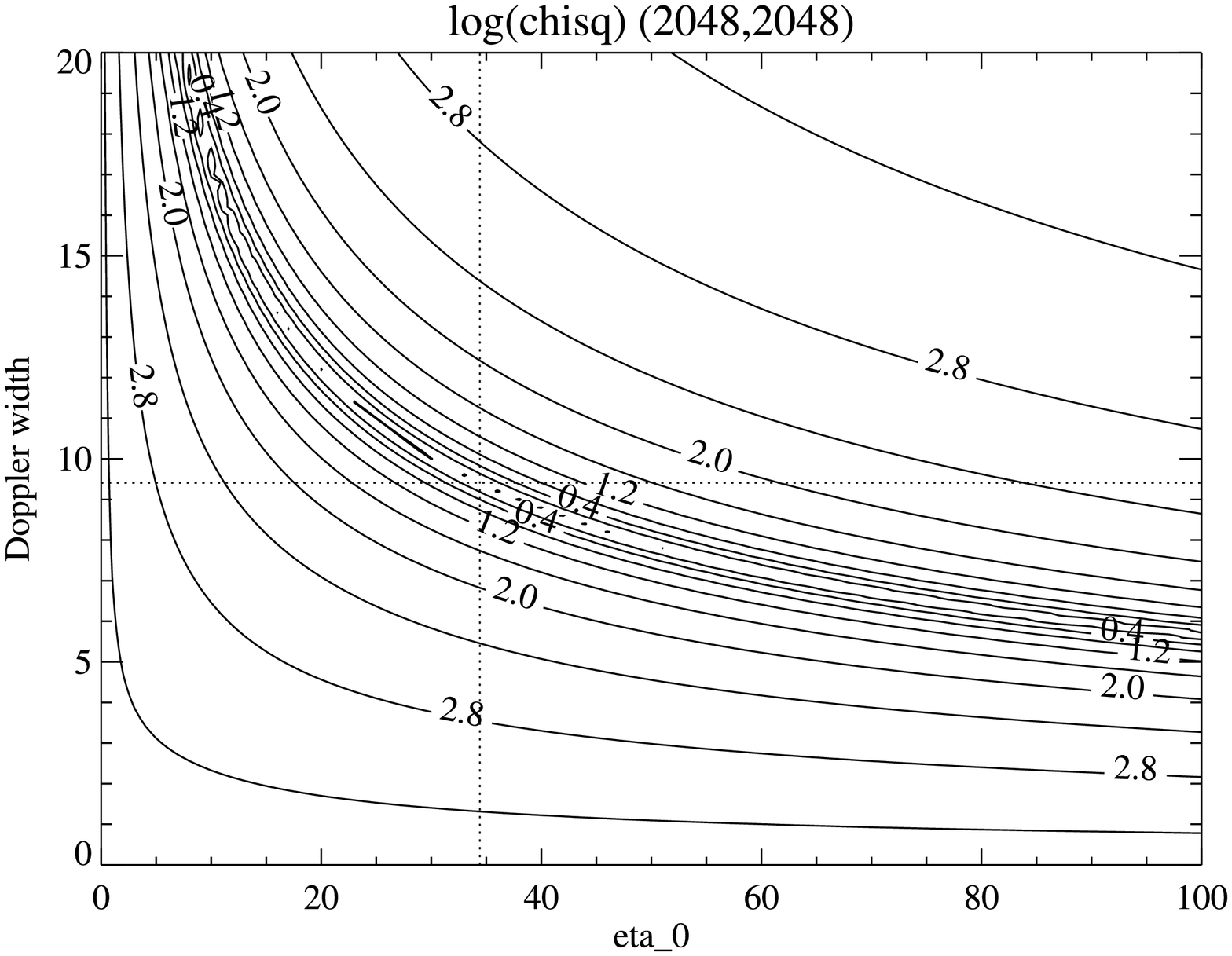}
\includegraphics[angle = 90,width=0.48\textwidth,bb=50 50  600 700, clip=]{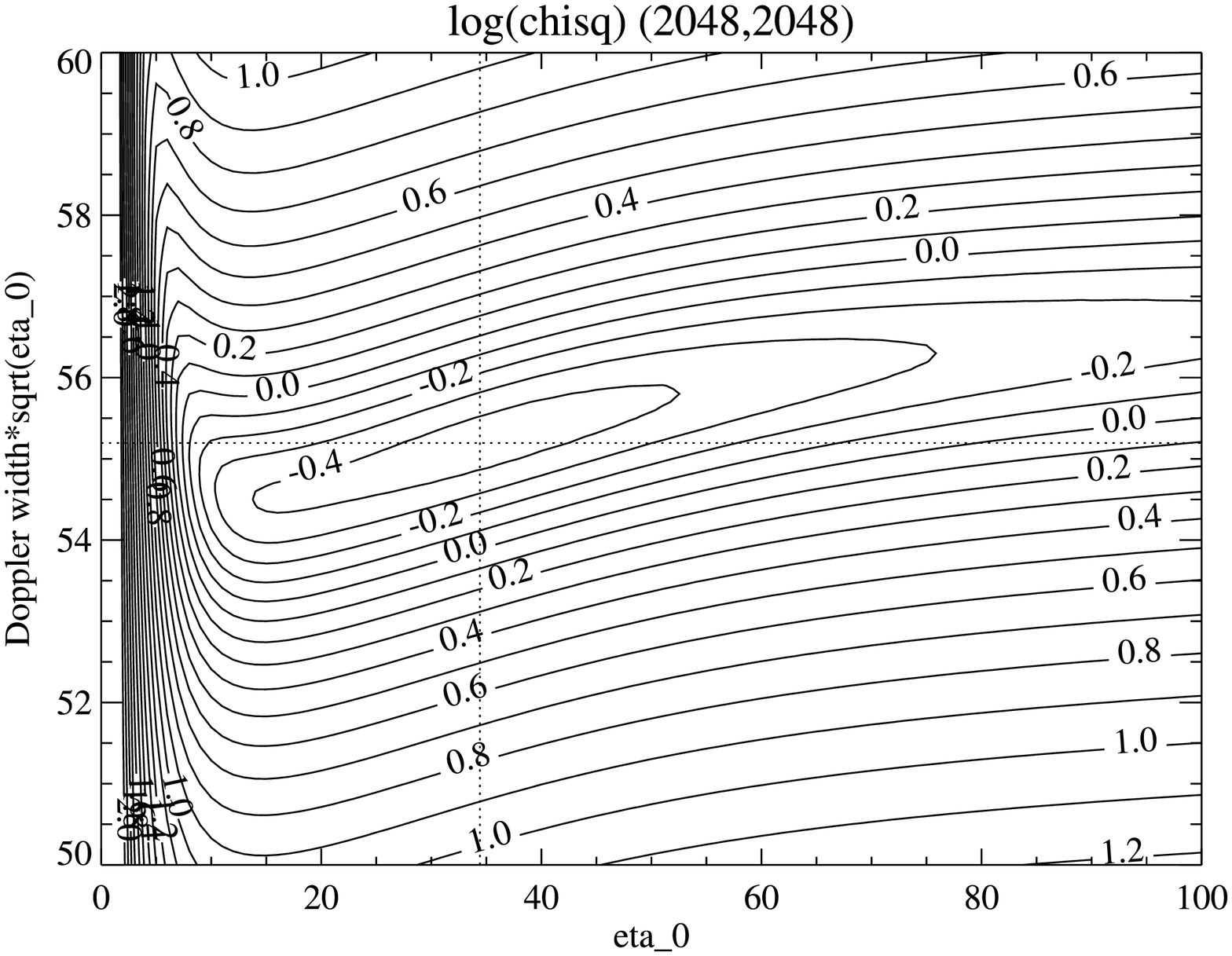}}
\caption{Contour plots of the $\chi^2$ surface as a function of two parameters. Left: logarithm of $\chi^2$ as a function of $\eta_0$ and $\Delta\lambda_{\rm D}$ for the inversion of a the central pixel of the HMI CCD ([2048,2048] in the CCD coordinates). Right: exact same plot represented, instead, as a function of $\eta_0$ and ($\Delta\lambda_{\rm D} \times \sqrt{\eta_0}$). The elliptical shapes of the $\chi^2$-contours in the second case show that the change of variable will help the algorithm localize the solution ({\it i.e.} the minimum $\chi^2$) faster.}
\label{fig:change_variable}
\end{figure}

In exploring the details of the evolution of $\chi^2$ during the inversion of selected pixels using the original VFISV code, it became apparent that the path to convergence was not always the fastest or the most efficient one. This stemmed, partly, from the degeneracy among the physical parameters in the model atmosphere.
For instance, there is a strong coupling between the terms of the source function, $B_0$ and $B_1$.
While the two parameters are difficult to determine individually,
their sum is tightly constrained by the continuum intensity.

While this is not, in general, a major problem for an algorithm using
the full Hessian, we nonetheless chose to fit for $B_0$ and $B_0+B_1$,
in order to make the inversion more well behaved.

When looking at the shape of the $\chi^2$ surface close to convergence as a function of different pairs of variables, the degeneracy problem becomes obvious. The left panel of Figure \ref{fig:change_variable} shows the logarithm of the $\chi^2$ surface as a function of $\eta_0$ on the {\it x}-axis and the Doppler width, $\Delta\lambda_{\rm D}$, on the {\it y}-axis.
The contours show a narrow {\em curved valley} of minimum $\chi^2$ values,
along which different combinations of the two parameters result in a very
similar goodness of the fit.
Such a curved valley tends to lead to very poor performance of the
LM algorithm.
Implementing a change of variable, that makes the contours of the
$\chi^2$ surface closer to elliptical (see right panel of the same figure),
leads to a $\chi^2$ surface that is better approximated by a quadratic form
in the parameters
(that is, has a larger range of validity of a Taylor expansion using 
the first and second partial derivatives)
and, in turn, to a much improved convergence of the 
the LM algorithm.
%to improve the solution in each iterative step \cite[see][]{press}.

Two changes of variable were implemented:

\begin{itemize}
\item $B_0$ and $B_1 \, \rightarrow \, B_0$ and $(B_0+B_1)$
\item $\eta_0$ and $\Delta\lambda_{\rm D} \, \rightarrow \, (\Delta\lambda_{\rm D} \times \sqrt{\eta_0 })$ and $\Delta\lambda_{\rm D}$
\end{itemize}

\noindent The variable changes are computed just before the matrix inversion and undone right after obtaining the improved model parameters (see a schematic chart of the LM algorithm in \cite{Borr11}). This means that the forward modeling and the definition of $\chi^2$ remain unchanged, and so does the physical model of the atmosphere.
By definition, these variable changes do not alter the position of the
minimum, expressed in the original variables, and so the position of the
global minimum is unchanged.
However, its path through the parameter space becomes more efficient after the variable change, which results in fewer iterations before convergence and hence a speed up of the inversion. The subsequent increase in speed achieved by the two variable changes described above results in a $\approx 10\%$ reduction in computing time for a full disk map in the HMI pipeline.

\subsection{Iteration Algorithm and Testing}

As part of the effort to speed up the inversions,
the iteration algorithm of VFISV was significantly improved.
Standard LM techniques are good at finding local minima in the minimization problem. However, they lack the capability to search the entire parameter space in pursuit of a global minimum. The approach that was followed for the \verb|fd10| version of the HMI pipeline inversion has two levels of
operation. The lower level takes an initial guess for the
parameters and uses a LM algorithm to find a
local minimum. The upper level tries to determine if the best solution found
so far is the global minimum, and if not, a new initial guess is provided
and the module calls the lower level again to try to find the global minimum; we call this a {\it restart}.

Several upper-level algorithms were tested against one another. In order to 
assess their performance, the quality of the fits obtained needed to be evaluated and compared to the time spent converging to the solution.
The selected time criterion was the average number of iterations
per pixel. While this is not exactly proportional to the running time, it
is a good proxy.
The criterion chosen to assess the quality of fits was 
the fraction of pixels for which the fitted $\chi^2$ was within a given
tolerance of that at the global minimum.

In order to evaluate the algorithms, a number of test datasets were selecthed that varied in complexity of the magnetic field and covered
a range of line-of-sight spacecraft-to-Sun velocities. Unfortunately, the solution of the inversion problem ({\it i.e.} the global minimum) for real observations of the Sun is unknown.  Therefore, the performance of the different algorithms had to be tested against some approximation to the solution.  

In order to find the best estimate of the global minimum, a fixed version of VFISV was used to invert the test dataset one hundred times, each time with a different initial guess that was chosen at random spanning the entire parameter space. For each inversion the code was forced to do two thousand iterations, allowing for restarts whenever the algorithm converged to a solution or failed to converge at all.   For each pixel, the solution with the lowest value of $\chi^2$ among the one hundred inversions was selected to represent the global minimum.   A composite set of inversion results was put together using the best solution for each pixel in the field of view. This was labelled as the {\em gold standard} (GS) and has an associated $\chi^2_{\rm GS}$ map.

The lower level of the algorithm is a fairly standard 
LM algorithm. The control parameter $\lambda$ is
lowered by a factor of 5 if $\chi^2$ decreases by more than a trivial
amount (0.0001), with a lower bound for the value of 0.0001,
and otherwise increased by an amount that depends on $\lambda$
(10 if $\lambda \le 0.01$ and 20 otherwise).
At each step, the parameter limits are enforced, as are limits on the
changes in some variables.
It should also be noted that the model derivatives are only calculated
when $\chi^2$ improves \citep{Borr11}, thereby saving a considerable time over
always calculating them.
The iteration stops if $\lambda$ exceeds a certain value (100) or if the
total number of iterations exceeds an upper bound (200). 
% or if several consecutive
%restarts arrive at the same solution independently of where they started in the parameter space.

At the end of each execution of the lower layer, the upper layer determines if
a new initial guess for the model parameters should be tried, in which case, it generates one.
A number of variations of the upper-level algorithm were thoroughly tested. 
They mostly differed in the way the new initial guess was chosen. 
The criteria for when a new guess and a subsequent restart is implemented were
determined in a semi-empirical manner and were tailored to the \verb|hmi.S_720s| data series. 
What follows is a description of the condition clauses for the upper-level algorithm
implemented in \verb|fd10| in order to find a global minimum in the inversion problem. This
particular set of conditions was chosen because it showed the best compromise between speed
and convergence to the {\em gold standard} \footnote{Note that the algorithm used to produce the {\em gold standard} was not a viable one for the HMI pipeline processing due to its high computation expense.}.

At the end of the first execution of the lower level code, the algorithm will likely exit the iteration loop and the inversion will spit out a solution. However, for the pixels that meet one of the following criteria, one single restart will be enforced:

\begin{itemize}
\item The field is low ($B < 30$\,G) or high ($B > 300$\,G).
\item The $\chi^2$ is larger than 20.
\item $\eta_0 < 5$ for $B < 300$\,G.
\item Points with large magnetic field inclinations or low values of the Doppler width ($\Delta\lambda_{\rm D} < 15$).
%\item Low field (B $<$ 300 G) points with $\eta_0 < 5$ and $\Delta\lambda_D > 20$
\end{itemize}

At the end of the second execution, an upper-level decision to implement further restarts will be made based on the following criteria:
\begin{itemize}
\item Low ($B < (35-N)$\,G) or high ($B > 500$\,G) fields , where $N$ is the number of restarts already performed.
\item Inclination within 30 degrees of vertical.
\item Very low field ($B < 45$\,G) points with $\eta_0 < 3.5$ and $\Delta\lambda_{\rm D} > 30$.
\end{itemize}

For the first and all subsequent odd restarts, a tailored guess
is made as to a good new starting point in the parameter space.
For the even numbered restarts, a random model perturbation is used.
In most cases, the new initial guess will converge to the same
solution as an earlier one. An additional control parameter is flagged when, 
after the first restart, the algorithm shows signs of heading inevitably to the 
same solution again, in which case it forces the exit of the iteration loop to avoid
unnecessary iterations. The algorithm will never allow the total number of iterations ({\it i.e.} the sum of the iterations performed in all the restarts) exceed the maximum allowed (200).

The combination of the LM with this upper-level algorithm ensures 
that the global minimum is almost always found without using an excessive amount of
computing time.
The statistics for VFISV \verb|fd10| code are as follows:

\begin{itemize}
\item Of all the pixels, $0.008\%$ have not converged to within 0.1 of the $\chi^2_{\rm GS}$ value and $0.185\%$ have not converged to within 0.01 of the $\chi^2_{\rm GS}$.
\item Of the pixels with $B>$ 300 G, $0.041 \%$ have not converged to within 0.1 and $0.097\%$ not to within 0.01 of $\chi^2_{\rm GS}$.
\item The rms error in $B$ is 3.00 G.
\item The average number of iterations is 28.22.
\end{itemize}

\subsection{Parameter Initialization}

The first initial guess of the atmospheric parameters for the inversion algorithm is mostly tailored to each pixel, although some of the parameters are given the same initial value for the entire field of view. This is true for the magnetic field strength, which is always initialized at 150\,G, and its LOS inclination, which is started at 90$^{\circ}$. However, other parameters are initialized differently for each individual pixel. 
A weak field approximation is used to calculate the initial value for the azimuth ($\psi = (1/2){\rm arctan}(Q/U)$), and a rough guess, based on the wavelength shift of the spectral line, is estimated for the velocity. The source function and its gradient are constrained to add up to the continuum intensity, and an estimate of the Doppler width ($\Delta\lambda_{\rm D}$) is obtained from the shape of the spectral line. However, these first guesses are not crucial because the restart system described above erases any memory 
of the initialization, especially for the pixels with a relatively large magnetic signal, for which restarts are enforced until the maximum allowed number of iterations is reached.

\section{Conclusion}\label{section:conclusions}

Some of the changes to the VFISV code reported in this paper were aimed at speeding up the code while having little or no impact on the accuracy of the solution found by the inversion algorithm. These include measures that prevent the algorithm from performing unnecessary calculations and from following inefficient paths through the parameter space in pursuit of the solution.
Certain modifications tried to palliate other performance issues that arise from the degeneracy among the atmospheric model parameters (which lead to multiple minima in the $\chi^2$) from the sometimes inadequate simplicity of the model chosen to represent the complexity of the real Sun, and from the data themselves, which have limited spectral and spatial resolution, are affected by photon noise, {\it etc}.

Throughout this manuscript we organized the changes into two classes: those that altered the shape of the $\chi^2$ function (and hence the model atmosphere), and those that did not. However, they are not all independent of one another, and they operate together to improve the overall efficiency and performance of VFISV in the HMI data pipeline.

\begin{itemize}
\item A set of weights for the Stokes profiles that provided the smoothest possible solution inside active regions was chosen. This renders less than optimal results in weakly magnetized areas where the polarization profiles are heavily dominated by the noise. Because of the smearing effect of the transmission filter profiles of the HMI instrument, the standard weighting system used to invert datasets from spectrograph type of instruments (such as ASP or {\it Hinode}/SOT-SP), does not work with HMI data. A custom set of weights was chosen and applied uniformly over the entire FOV.

\item A regularization term that penalizes high values of $\eta_0$ was added to the merit function. This mitigates the double minima problem in the parameter space and helps prevent the code from converging to unphysical solutions.

\item Limits on the range of each atmospheric parameter were set to prevent the algorithm from probing extreme unphysical values.

\item A diagnostic procedure  that identifies possibly problematic pixels enforces a series of inversion restarts that attempt to localize the global minimum. Coupled to this, a discriminating convergence criteria ensures a very accurate solution while keeping the average number of iterations relatively low ($\approx 30$).

\item A major speed improvement (about 64\% reduction in computing time) was achieved, without compromising the accuracy of the solution, by considering a spectral range of $\pm 2$ \AA\ around the central wavelength but limiting the explicit forward modeling calculation to an inner $\pm 0.65$ \AA\ range. This hybrid approach accounts for the light that passes through the secondary lobes of the HMI transmission filter profiles without doing the detailed spectral line calculation far out into the continuum.

\item An additional 10\% increase in speed was obtained after performing two changes of variable on the atmospheric model parameters before solving the linear system of equations at each iterative step. This modification prevented the code from performing useless iterations in the parameter space by helping it find a more direct route to the solution. 

\item Coding errors were identified and corrected. Some of these are reported in the Appendix.
\end{itemize}

%\acknowledgments
\begin{acks}
Much of the reported effort
was supported by Stanford University NASA Grant NAS5-02139
for SDO/HMI commissioning and pipeline code implementation; KDL and GB
also acknowledge PO\# NNG12PP28D/C\# GS-23F-0197P from NASA/Goddard Space Flight Center.
The National Center for Atmospheric Research is sponsored by the National Science Foundation.
\end{acks}

\appendix %[Coding Errors]
\label{section:bugs}

This section reports some coding mistakes that have been spotted in VFISV since its original release and that have been fixed in the \verb|fd10| HMI data pipeline version. This version can be found through the JSOC (Joint Science Operations Center) CVS tree at Stanford.

\section{Voigt Function}

The Voigt function is a computationally intensive part of the forward modeling of the spectral line. In order to speed up the calculation, the actual Voigt function is only computed for small parts of the spectral range, close to the line core and far into the wings. In the intermediate wavelength regime the Voigt function is approximated by a Taylor expansion using a lookup table for the Taylor coefficients as a function of the central wavelength of the Voigt profile. Two bugs were found in the coding of the Voigt function of the spectral line.

An error in the wavelength indexing for the lookup table (inside the \verb|voigt_| \verb|taylor.f90| routine) resulted in sudden unphysical changes in the integrated area of the Voigt profile as a function of the central wavelength. This led to a periodic roughness pattern of the $\chi^2$-surface that resulted in the appearance of multiple local minima along the magnetic field direction when the algorithm was approaching the solution. Instead of being evenly spread (due to the effect of the photon noise), the solutions would preferentially fall along ridges with an 11 gauss periodicity in the magnetic field. 

A second coding mishap in the calculation of the Voigt profile was a missing factor of 2 in the computation of the imaginary part of the Taylor expansion of the Voigt function, which affected the magneto-optical effects in strong field regions. This error is heritage of the same missing factor of 2 in a FORTRAN routine (\verb|voigt.f|, originally developed by Harvey and Nordlund), that calculates the Voigt function. This routine has been used in a number of spectral line inversion codes for the calculation of the line profiles. 
Albeit small, the effect on the synthetic Stokes profiles lead to inhomogeneous patterns in the inverted magnetic fields inside sunspot umbrae.

\section{Uncertainties}

The photon noise in the observed Stokes profiles propagates to the atmospheric model inferred by the inversion. When the algorithm has converged and the optimal model has been found, VFISV computes the formal uncertainties corresponding to each physical parameter.
The variance associated with the retrieved model parameter $a_j$ is taken to be proportional to the corresponding diagonal element of the inverse of the curvature matrix $\alpha_{jj}^{-1}$ \cite[see, for instance,][]{bevington1969}:

\begin{equation}
\sigma^2_{a_j} = \frac{\chi^2}{ N_{{\rm FREE}}} \alpha^{-1}_{jj} 
\end{equation}

\noindent where the curvature matrix is the second cross-partial derivative of $\chi^2$ with respect to combinations of two of the model parameters:

\begin{equation}
\alpha_{jk} = \frac{1}{2} \frac{\partial^2\chi^2}{\partial a_j \partial a_k }.
\end{equation}

The non-diagonal elements of the inverse of the curvature matrix, ${\bf \alpha}^{-1}_{jk}$, are proportional to the covariances, $\sigma^2_{jk}$, between each pair, $a_j$ and $a_k$. These elements give a sense of the degeneracy between any two given parameters of the model atmosphere.

An index switch inside the nested loop that calculates the elements of the covariance matrix (subroutine \verb|get_err| of the \verb|inv_utils.f90| module) led to huge uncertainty values that were many orders of magnitude too large. After correcting for this problem, the magnitudes of these formal uncertainties lie within reasonable values and compare to those obtained through a Monte Carlo experiment.

\end{article}

\end{document}